\documentclass[prd,showkeys,preprintnumbers,floatfix,superscriptaddress, nofootinbib]{revtex4}
\usepackage{float}
\usepackage{nicefrac}
\usepackage{slashed}
\usepackage{mathtools}
\usepackage{amsfonts} 
\usepackage{amssymb} 
\usepackage{amsmath} 
\usepackage{esint}
\usepackage{graphicx} 
\usepackage{subfigure} 
\usepackage{array} 
\usepackage{dcolumn} 
\usepackage{bm} 
\usepackage{latexsym} 
\usepackage{longtable} 
\usepackage{hyperref} 
\usepackage{verbatim}
\usepackage{epsfig}
\usepackage[usenames,dvipsnames,svgnames,table]{xcolor}
\newcommand{\ximn}{\mathbf{\xi}_{\mathbf{mn}}}

\begin{document}
  \preprint{\vbox{\hbox{CERN-TH-2018-109} }}
   \preprint{\vbox{\hbox{INT-PUB-18-019} }}
 \preprint{\vbox{\hbox{JLAB-THY-18-2697} }}
\title{Finite-volume effects due to spatially nonlocal operators}
\author{Ra\'ul A. Brice\~no}
\email[e-mail: ]{rbriceno@jlab.org}
\affiliation{Thomas Jefferson National Accelerator Facility, 12000 Jefferson Avenue, Newport News, Virginia 23606, USA}
\affiliation{ Department of Physics, Old Dominion University, Norfolk, Virginia 23529, USA}

\author{Juan V. Guerrero}
\email[e-mail: ]{juanvg@jlab.org}
\affiliation{Thomas Jefferson National Accelerator Facility, 12000 Jefferson Avenue, Newport News, Virginia 23606, USA}
\affiliation{Hampton University, Hampton, Virginia 23668, USA}

\author{Maxwell T. Hansen}
\email[e-mail: ]{maxwell.hansen@cern.ch}
\affiliation{Theoretical Physics Department, CERN, 1211 Geneva 23, Switzerland}

\author{Christopher J.~Monahan}
\email[e-mail: ]{cjm373@uw.edu}
\affiliation{Institute for Nuclear Theory, University of Washington, Seattle, Washington 98195, USA}

\date{\today}

\begin{abstract}

Spatially nonlocal matrix elements are useful lattice-QCD observables in a variety of contexts, for example in determining hadron structure. To quote credible estimates of the systematic uncertainties in these calculations, one must understand, among other things, the size of the finite-volume effects when such matrix elements are extracted from numerical lattice calculations. In this work, we estimate finite-volume effects for matrix elements of nonlocal operators, composed of two currents displaced in a spatial direction by a distance $\xi$. We find that the finite-volume corrections depend on the details of the matrix element. If the external state is the lightest degree of freedom in the theory, e.g.~the pion in QCD, then the volume corrections scale as $ e^{-m_\pi (L- \xi)} $, where $m_\pi$ is the mass of the light state. For heavier external states the usual $e^{- m_\pi L}$ form is recovered, but with a polynomial prefactor of the form $L^m/|L - \xi|^n$ that can lead to enhanced volume effects. These observations are potentially relevant to a wide variety of observables being studied using lattice QCD, including parton distribution functions, double-beta-decay and Compton-scattering matrix elements, and long-range weak matrix elements.
 \end{abstract}

\keywords{parton distribution functions, lattice QCD, finite volume}

\nopagebreak
\maketitle

\section{Introduction}

One of the fundamental goals in theoretical nuclear physics is the prediction of hadron structure from first-principles calculations based on the underlying gauge theory of the strong nuclear force, quantum chromodynamics (QCD). Quarks and gluons, the degrees of freedom of QCD, are confined into color-singlet bound states that are observed experimentally. The internal structure of these hadrons, encoded in the spatial, momentum and angular-momentum distributions of the constituent quarks and gluons, is inherently nonperturbative and poorly understood. Forthcoming data from the 12 GeV upgrade at JLab \cite{Dudek:2012vr} and a future electron-ion collider \cite{Accardi:2012qut} will provide experimental insight into hadron structure, but a complete understanding of the experimental data requires a concomitant improvement in our theoretical understanding.

Observables related to hadronic structure are most naturally formulated using light-cone quantization, and this provides a serious challenge for lattice QCD, which is necessarily restricted to Euclidean-signature correlation functions. As a specific example, parton distribution functions (PDFs), which capture the distribution of the longitudinal momentum of a hadron among its constituent quarks and gluons, are defined via lightlike separated fields and thus cannot be directly accessed in a Euclidean spacetime, where $x^2=0$ defines a single point rather than a cone. In the past two decades, various ideas have been proposed to overcome this challenge, and thereby calculate PDFs and similar observables from lattice QCD  \cite{Aglietti:1998ur,Abada:2001if,Detmold:2005gg,Braun:2007wv,Ji:2013dva,Radyushkin:2016hsy,Radyushkin:2017cyf,Chambers:2017dov,Hansen:2017mnd}.

Although the details differ, these methods generally require the evaluation of matrix elements of nonlocal operators, frequently using hadronic states with high momentum. Preliminary results for several of these ideas have now appeared \cite{Lin:2014zya, Alexandrou:2015rja,Chen:2016utp, Alexandrou:2016jqi, Lin:2017ani, Chen:2017mzz, Orginos:2017kos, Chambers:2017dov, Ma:2017pxb, Bali:2017gfr, Alexandrou:2018pbm} (see Ref.~\cite{Lin:2017snn} for a recent review), but an understanding of all systematic uncertainties is not yet feasible. In general, the systematic uncertainties associated with such calculations include discretization effects, which may be significant for high-momentum states, uncertainties associated with the momentum of the hadron \cite{Chen:2017lnm,Bali:2016lva}, truncation errors arising from perturbative renormalization or matching \cite{Constantinou:2017sej,Alexandrou:2017qpu,Radyushkin:2017ffo}, and finite-volume effects. 
 
In addition to systematically reducing computational uncertainties, it is crucial to shore up the theoretical foundations of the approach. Here, significant progress has been made in understanding issues associated with the renormalization of the Wilson-line operator \cite{Ishikawa:2016znu,Monahan:2016bvm,Ji:2017oey,Ishikawa:2017faj,Green:2017xeu,Alexandrou:2017huk,Chen:2017mie,Chen:2017mzz}, the existence of factorization theorems \cite{Zhang:2018ggy,Izubuchi:2018srq}, and the role of the Euclidean signature in lattice calculations \cite{Briceno:2017cpo}.

In this work we study the finite-volume artifacts associated with spatially displaced currents. We do so by perturbatively studying a matrix element of a product of two currents in a toy theory with one light degree of freedom (corresponding to the pion in QCD) and one heavy (corresponding to a nucleon or heavy meson) degree of freedom. Proposals to extract hadronic structure observables from lattice-QCD calculations using these types of operators, in place of those defined with a  Wilson line, appeared in Refs.~\cite{Ma:2017pxb, Braun:2007wv}, with the first numerical results for pion distribution amplitudes presented in Ref.~\cite{Bali:2017gfr}. 

To perform this type of finite-volume analysis for Wilson-line-based operators would require a low-energy representation for these, more complicated, nonlocal objects~\cite{Chen:2001nb, Kivel:2002ia, Kivel:2007jj,Diehl:2005rn}. One possible avenue is to build an effective field theory based on the method of Ref.~\cite{Green:2017xeu}, in which an auxiliary heavy-quark field enables one to write gauge-invariant extended operators in terms of products of quark bilinears. However, this goes beyond the scope of the present work, and we focus our attention on composite bilinear currents, the hadronic representation of which is more straightforward.

For lattice calculations of hadronic masses, and other properties defined through local operators, finite-volume effects lead to corrections of the form $\mathcal{O}(e^{-m_\pi L})$~\cite{Luscher:1985dn, Colangelo:2002hy, Colangelo:2003hf,Beane:2004tw, Colangelo:2004xr,Beane:2004rf, Bijnens:2013doa}, where $L$ is the linear extent of the cubic spatial volume.%
\footnote{More precisely, Ref.~\cite{Luscher:1985dn} found that the leading exponential correcting a stable particle mass is $e^{- \sqrt{3} m_\pi L /2}$ in the case of odd-legged interaction vertices and $e^{- m_\pi L}$ for theories with a $\mathbb Z_2$ symmetry.} 
Numerical data are expected to be described by this leading exponential form, with a power-law prefactor, provided one performs the calculation with asymptotically large volumes, $m_\pi L \gg 1$. (In practice, $m_{\pi} L \gtrsim  4$ is generally sufficient.) However, in the presence of a second infrared (IR) length scale, such as the current separation in a spatially extended operator, one naturally expects the finite-volume effects to be modified.

For matrix elements of composite currents, we show that finite-volume effects take the form
\begin{equation}
 \langle   M \vert  \mathcal J(0, \pmb \xi) \mathcal J(0)      \vert    M  \rangle_L -  \langle   M \vert  \mathcal J(0, \pmb \xi) \mathcal J(0)      \vert   M \rangle_\infty  =  P_a(\pmb \xi, L)    e^{- M (L-  \xi) } +  P_b(\pmb \xi, L)    e^{- m_\pi L  }  + \cdots \,,
\label{eq:main_result}
\end{equation} 
where the left-hand side represents the difference between the finite-volume matrix element (obtained via lattice QCD) and its infinite-volume limit. The external states here are zero-momentum, single-particle states, labeled by their physical mass, $M$; $\pmb \xi$ is the displacement vector within the composite current, and $\xi = \vert \pmb \xi \vert$ is its magnitude. To derive this result, we assume $m_\pi L \gg m_\pi \xi \gtrsim  1$.

The right-hand side of Eq.~(\ref{eq:main_result}) gives the leading finite-volume effects. We focus on two terms, one scaling with the mass of the external state and the other with the mass of the lightest degree of freedom. In the case where these two coincide, the first term scales as $e^{- m_\pi (L - \xi)}$ and is expected to dominate the volume effects once $\xi$ becomes a non-negligible fraction of $L$. By contrast, if $M \gg m_\pi$, as in the case of a nucleon or heavy meson, then the second term dominates. Both terms have polynomial prefactors, denoted $P_a$ and $P_b$, with terms scaling as $L^m/|L - \xi|^n$. Such factors can also have a significant impact on the size of finite-volume corrections if $\xi$ is non-negligible compared to the box size. Finally, the ellipsis in Eq.~(\ref{eq:main_result}) represents subleading exponentials.

\begin{figure}[t]
\begin{center}
\includegraphics[width=\textwidth]{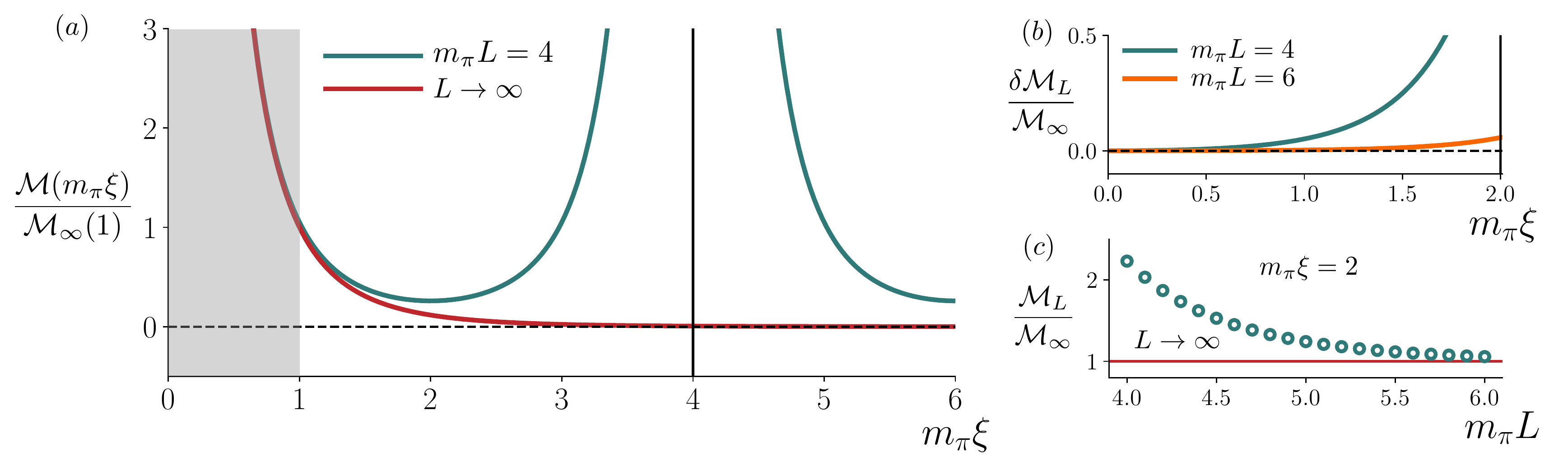}
\caption{Finite- versus infinite-volume behavior of nonlocal matrix elements. These plots were made using the tree-level result derived in the next section with pions as the external states. Subfigure $(a)$ shows how, as $\xi$ is varied, the finite-volume matrix element develops large deviations from its infinite-volume counterpart. For $m_\pi\xi \lesssim 1$, indicated by the shaded region, high-energy scales are sample so that the effective field theory is expected to break down. Subfigure $(b)$ shows the fractional difference between finite- and infinite-volume matrix elements, $\vert \mathcal M_L - \mathcal M_\infty \vert / \vert \mathcal M_\infty \vert$. Finally, (c) shows the finite-volume matrix element, $\mathcal M_L$, as a function of $L$, together with its infinite-volume limit. For fixed $\xi$, finite-volume effects for $\mathcal M_L$ decay with $L$ as $e^{- m_\pi L}$. These are enhanced by a $e^{ m_\pi\xi}$ prefactor relative to the typical, $\mathcal O(1) \times e^{-m_\pi L}$, finite-volume effects for local operators.
\label{fig:periodic} }
\end{center}
\end{figure}

To better understand these volume effects we note that, while the infinite-volume matrix element generally decays as a function of $\xi$, its finite-volume counterpart must be periodic, with periodicity $L$. Thus, as we illustrate in Fig.~\ref{fig:periodic}(a), the differences between the finite- and infinite-volume objects become arbitrarily large as $\xi$ approaches $L$. We are interested in the onset of this effect for $\xi \ll L$. As we show in Fig.~\ref{fig:periodic}(b), in the case where the external state is light, the finite-volume effects grow exponentially with $\xi$, exhibiting $\sim10 \%$ deviations for $\xi \sim L/4$ when $m_\pi L=4$. For this same volume, Fig.~\ref{fig:periodic}(c) shows that these volume effects can be removed by fitting to a decaying exponential in $L$ at fixed $\xi$. We stress that the details of these features hold only for matrix elements built from products of local currents.

The origin of periodicity for matrix elements built from products of local operators is straightforward: Given that the quark fields and the gauge links are periodic in all spatial directions, the same must be true for any local current $\mathcal J$ constructed from these fields and links. The periodicity property $\mathcal J(t, \textbf x) = \mathcal J(t, \textbf x + L \textbf e_i)$, with $\textbf e_i$ a unit vector in the $x$, $y$, or $z$ direction, is then directly inherited by matrix elements constructed from products of such currents at different locations.

However, this argument does not hold for nonlocal quark bilinears connected by Wilson lines, i.e.~the type of nonlocal operator used to extract quasi- and pseudo-PDFs. Defining $W[x + \xi \textbf e_i ,x]$ as the straight Wilson line connecting the points $x + \xi \textbf e_i$ and $x$, given by
\begin{align}
W[x + \xi \textbf e_i ,x] \equiv U_i( x + (\xi-a) \textbf e_i)\, U_i( x + (\xi-2 a) \textbf e_i)  \times \cdots \times  U_i( x ) \,,
\end{align}
one can construct a gauge-invariant nonlocal operator by contracting this with quark and antiquark fields at $x$ and $x + \xi \textbf e_i$, respectively. The quark fields and the gauge fields are periodic in the spatial directions, but for fixed $\textbf x$, there is no periodicity in the coordinate $\xi$. In particular, wrapping around the torus $n$ times gives
\begin{equation}
\overline{q} \big (x + (\xi + n L ) \textbf e_i \big) \, W \big [x + (\xi + n L) \textbf e_i ,x \big] \,{q} \big (x \big) =    \overline{q}(x + \xi   \textbf e_i)  \, W[x + \xi   \textbf e_i ,x] \, \Big ( W[x + L   \textbf e_i ,x]^n \Big )   \,{q}(x)  \,,
\end{equation}
where the factor in parentheses on the right-hand side breaks the naive periodicity relation, $\mathcal O (\xi + L) = \mathcal O(\xi)$. 

On the one hand, this additional factor may lead to matrix elements of this operator being closer to their infinite-volume counterparts than in the case of products of currents that satisfy $\xi$-periodicity. On the other hand, the fact that the boundary conditions are felt by both the quark fields and individual links leads us to expect that large values of $\xi$ may generate enhanced volume effects in this case as well. 

As we describe in detail in Sec.~\ref{sec:toy_model} below, we derive Eq.~(\ref{eq:main_result}) and Fig.~\ref{fig:periodic} using a toy theory with two relativistic scalar particles. A rigorous demonstration that the scaling also holds in QCD would require first defining a specific matrix element, then developing a low-energy effective-field-theory description (based in chiral perturbation theory) and finally calculating finite-volume corrections. However, since our result only relies on the appearance of scalar propagators with the light particle mass, together with the scale $\xi$ that characterizes the operator nonlocality, we expect that a more realistic description would change only the detailed form of $P_a(\pmb \xi, L) $ and $P_b(\pmb \xi, L) $,  and not the overall exponential behavior. 

To close the Introduction, we comment on a number of other examples in which finite-volume effects on nonlocal matrix elements have already been discussed in the literature.

The authors of Ref.~\cite{Christ:2015pwa} describe formalism for removing finite-volume effects in neutral kaon mixing. The starting point is a matrix element reminiscent of that considered here, defined with external kaon states and two insertions of the weak Hamiltonian. In contrast to the matrix elements in this study, however, the currents are also separated in Euclidean time. By summing over time slices, the authors demonstrate how to identify a finite-volume version of $\Delta M_K$. In a second step, the leading finite-volume effects are removed using a generalization of the Lellouch-L\"uscher formalism \cite{Lellouch:2000pv}. The step of identifying the finite-volume version of $\Delta M_K$ relies on picking out a single term in the temporally summed correlator. This term in isolation has power-law volume effects associated with on-shell intermediate states, i.e.~effects parametrically larger than those identified in the present study. It is these volume artifacts that are corrected via the extended Lellouch-L\"uscher formalism.

In a different application, in Ref.~\cite{Hansen:2017mnd}, one of us considered an approach for extracting total decay and transition rates from temporally displaced currents with single-particle external states. The method requires estimating a smeared-out inversion of the Laplace transform, for example by using the Backus-Gilbert method \cite{Backus1,Backus2,Press:2007zz}. As discussed in detail in Ref.~\cite{Hansen:2017mnd}, this smearing suppresses finite-volume effects in the target observable. The infinite-volume observable must then be extracted by identifying an optimal trajectory in the two-coordinate plane of box size, $L$, and smearing width $\Delta$. The enhanced volume effects identified here will likely influence this optimal trajectory, but the detailed consequences are not clear and are the subject of future work.

The remainder of this article is organized as follows. In Sec.~\ref{sec:toy_model}, we explain the setup of our calculation, including the detailed definition of the toy theory and the external currents. We then summarize the general framework for calculating finite-volume effects in nonlocal matrix elements and apply this to the tree-level diagram of Fig.~\ref{fig:rules_LO}(b). In Sec.~\ref{sec:heavy}, we provide a detailed study of the finite-volume effects in one-loop diagrams, focusing on the case where the external state in the matrix element is a heavy particle, for example a nucleon. We then extract the large-volume scaling of these diagrams in Sec.~\ref{sec:asymptotics}, and deduce the result summarized by Eq.~(\ref{eq:main_result}). In Sec.~\ref{sec:summary}, we briefly conclude and outline possible future work. Technical details of certain functions used in the analysis are discussed in the Appendix.

\section{Setup and simple example}

\label{sec:toy_model}

\begin{figure}[t]
\centering
   \subfigure[]{\includegraphics[width =1\textwidth]{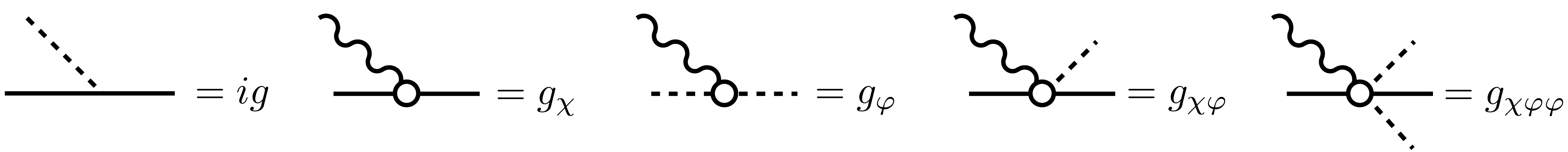}}
   \subfigure[]{\includegraphics[width =.3\textwidth]{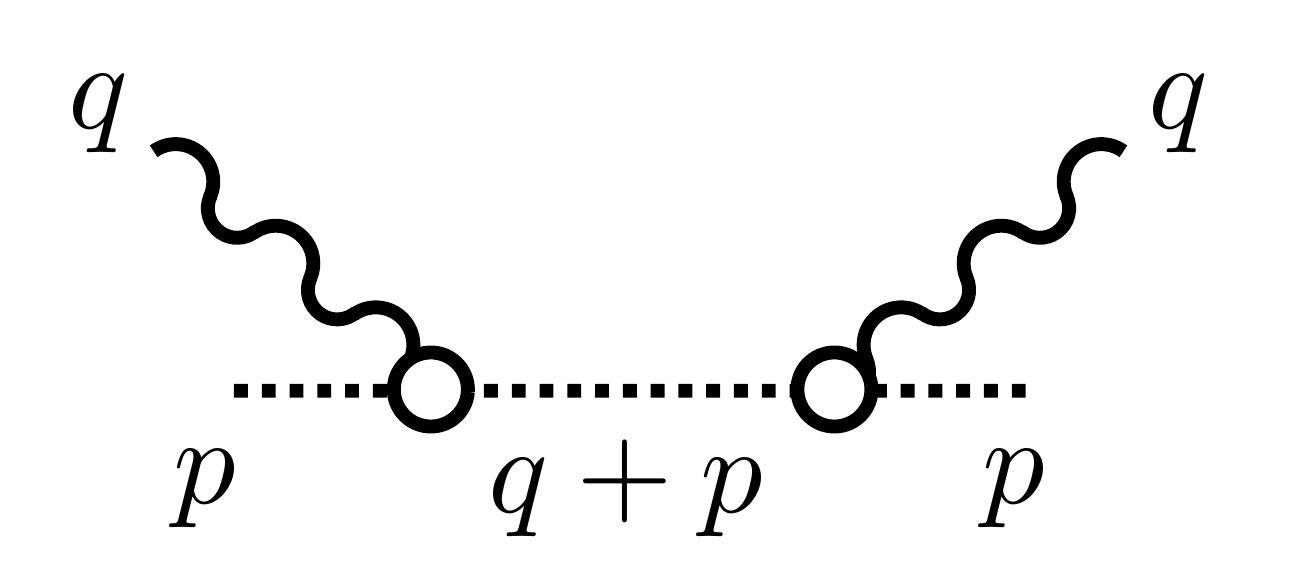}}
   \subfigure[]{\includegraphics[width =.2\textwidth]{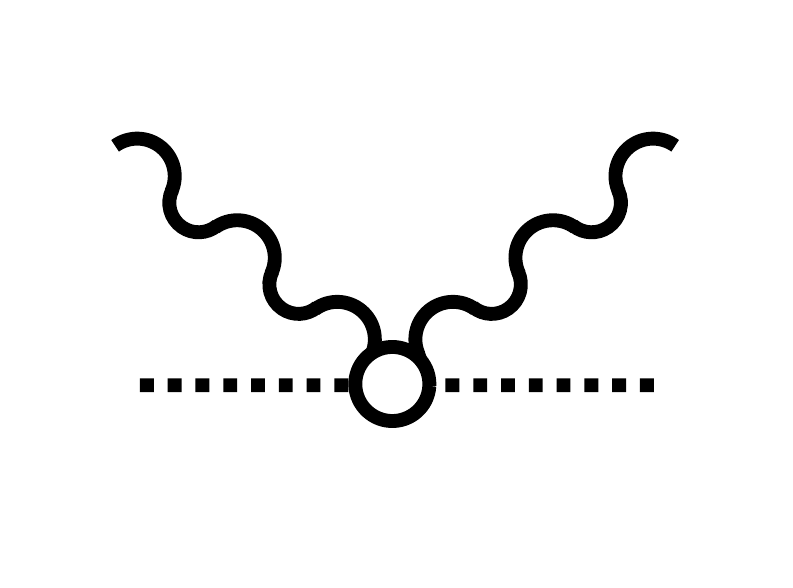}}
\caption{(a) Feynman rules for the EFT described in Sec.~\ref{sec:toy_model}. The dashed lines denote the lighter particle, $\varphi$, and the solid lines denote the heavier particle, $\chi$. (b) The leading-order contribution to the matrix element ${\cal M}(\pmb \xi, \textbf p)$ with $\varphi$ external states. (c) Contact interactions that may arise in EFTs. 
}\label{fig:rules_LO} 
\end{figure}

To determine finite-volume corrections to field-theoretic observables analytically, it is often useful to work with a low-energy effective theory (EFT), e.g.~chiral-perturbation theory ($\chi$PT) for QCD. This is natural because quarks are confined, and only the hadrons (the degrees of freedom in $\chi$PT) can propagate long distances and feel the finite-volume boundary conditions.

As a first step, in this work we study a toy theory that is expected to capture the basic scaling of the finite-volume corrections. We present a formal expression [Eq.~(\ref{eq:master}) below] that gives the finite-volume effects of spatially nonlocal currents from an arbitrary Feynman diagram. Using this result, we determine the finite-volume corrections from the tree-level diagram shown in Fig.~\ref{fig:rules_LO}(b) and, in the following two sections, consider the role of one-loop corrections in volume effects. 

We consider a theory with two scalar particles, $\varphi$ and $\chi$, with physical pole masses satisfying $m_{\varphi} < m_{\chi}$. Here $\varphi$ plays the role of the pion in QCD and $\chi$ that of the nucleon or a heavy meson. Using baryon and heavy meson $\chi$PT~\cite{Weinberg:1991um,Weinberg:1990rz, Wise:1992hn} as an inspiration, these states couple via a $\varphi \chi^2$ vertex of which the Feynman rule is given in Fig.~\ref{fig:rules_LO}(a). In QCD, pions are pseudo-Goldstone bosons and thus have derivative couplings to heavier particles. However, the exact form of these couplings does not change the leading exponential behavior of the theory, so here we only consider a momentum-independent coupling, labeled $g$. Similarly, we restrict attention to scalars, as details associated with spin and isospin are not expected to change the overall scaling of the finite-volume corrections.

The $\varphi$ and $\chi$ states couple to a renormalized external current, given by
\begin{equation}
\mathcal J(x) = \frac12  Z_{\varphi } g_{\varphi} \varphi^2 + \frac12 Z_{\chi } g_{\chi} \chi^2 + \frac12  Z_{\chi \varphi} g_{\chi \varphi} \chi^2 \varphi + \frac14  Z_{\chi \varphi\varphi} g_{\chi \varphi\varphi} \chi^2 \varphi^2 +\cdots \,,
\label{eq:currentdef}
\end{equation}
that generates the additional Feynman rules shown in Fig.~\ref{fig:rules_LO}(a).

The renormalization factors $Z_{\varphi }$ and $Z_{\chi }$ are inherited from the mass terms in the Lagrangian (with the scheme fixed by $\mathcal L \supset (1/2) m_{\chi}^2 Z_{\chi }  \chi^2$). The three-point renormalization is inherited in a similar way from $ \mathcal L \supset  (1/2) g Z_{\chi \varphi} \chi^2 \varphi$, with the scheme that the amputated three-point function equals $i g$ (its tree-level value) when all $p^2 = 0$.  A similar scheme can be used for the $\chi^2 \varphi^2$ term, although the coupling does not appear in the Lagrangian.\footnote{One possible approach is to include such a coupling, $ \mathcal L \supset   (1/4) \lambda Z_{\chi \varphi \varphi} \chi^2 \varphi^2$; define $Z_{\chi \varphi \varphi}$ such that $i \lambda$ coincides with the amputated, one-particle irreducible four-point function at $p^2=0$; and then take the $\lambda \to 0$ limit.} Finally, the ellipsis in Eq.~(\ref{eq:currentdef}) stands for terms with higher orders in $\varphi$ and $\chi$. 

The final step is to define a power-counting scheme for the theory. We take $g \sim g_\varphi \sim g_\chi \sim g_{\varphi \chi}/g$. As we consider matrix elements with two insertions of the local current, leading-order (LO) contributions scale as $g_{\varphi}^2 \sim g_{\chi}^2$ and next-to-leading-order (NLO) as $g_{\varphi}^2 g^2 \sim g_{\chi}^2 g^2 \sim g_{\varphi \chi}^2$.

We are now ready to set up our general approach for determining finite-volume effects in Feynman diagrams contributing to matrix elements of spatially nonlocal operators. We define the infinite-volume matrix element as
\begin{equation}
\mathcal M^{}_{\infty}(\pmb \xi, \textbf p)  \equiv \langle \textbf p \vert  \mathcal J(0, \pmb \xi) \mathcal J(0)    \vert \textbf p \rangle \,,
\end{equation}
where $\vert \textbf p \rangle$ is a single-particle state with momentum $\textbf p$, either a $\varphi$ or a $\chi$ to be specified below. Now note that any diagram, $d$, contributing to this quantity can be written as
\begin{equation}
\label{eq:Minfgen}
\mathcal M^{(d)}_{\infty}(\pmb \xi, \textbf p) = \int_q e^{i \textbf q \cdot \pmb \xi} \int_{k_1} \cdots \int_{k_{n-1}} (-i)^{n} D^{(d)}(p,q, k_1, \cdots, k_n) \,,
\end{equation}
where we have introduced the shorthand
\begin{equation}
\int_q \equiv \int \frac{d^4 q }{(2 \pi)^4} \,.
\end{equation}
In Eq.~(\ref{eq:Minfgen}), $(-i)^{n}D(\cdots )$ is the standard integrand, constructed according to the usual Feynman rules, and containing all couplings and symmetry factors. The separation of the $(-i)^{n}$ factor simplifies the relation to Euclidean-signature quantities. In particular, from the Wick rotation we find
\begin{equation}
\mathcal M^{(d)}_{\infty}(\pmb \xi, \textbf p) =  \int_{q_E} e^{i \textbf q \cdot \pmb \xi} \int_{k_{1,E}} \cdots \int_{k_{n-1,E}}   D^{(d)}_E(p_E,q_E, k_{1,E}, \cdots, k_{n,E}) \,,
\end{equation}
where $D^{(d)}_E(p_E,q_E, k_{1,E}, \cdots, k_{n,E}) \equiv D^{(d)}(p,q, k_1, \cdots, k_n)$ is the usual integrand that one would construct with Euclidean Feynman rules. 

As an example, for the leading-order diagram shown in Fig.~\ref{fig:rules_LO}(b), the Minkowski integrand is
\begin{equation}
D^{(\text{LO})}(p,q) = \frac{1}{(-i)}  g_\varphi^2 \frac{i}{(q+p)^2-m^2_\varphi+i\epsilon} = \frac{g_\varphi^2}{-(p+q)^2 + m_\varphi^2 - i \epsilon} \,,
\end{equation}
and Wick rotation gives
\begin{equation}
D^{(\text{LO})}_E(p_E,q_E) = \frac{g_\varphi^2}{(p_E+q_E)^2 + m_\varphi^2 } \,,
\end{equation}
consistent with the usual Feynman rules. Each loop introduces a factor of $( -i)$ to the Minkowski integrand, but in our convention this is factored out to preserve $D$ as defined in the two signatures.\footnote{Of course, for the final quantity we have no freedom in the convention. The Wick rotation preserves the value of $\mathcal M^{(d)}_{\infty}(\pmb \xi, \textbf p)$ by construction. But this correspondence is spoiled in the integrands by factors of $i$ that cancel with $q^0 = i q_4$. Our definition of $D$ simply compensates these factors to give $D = D_E$.}

We now give our general expression for the finite-volume effects from spatially nonlocal currents. From the Poisson summation formula, it follows that the finite-volume residue for any given diagram can be written as
\begin{align}
\delta \mathcal M^{(d)}_{L}(\pmb \xi, \textbf p) & \equiv \mathcal M^{(d)}_{L}(\pmb \xi, \textbf p) - \mathcal M^{(d)}_{\infty}(\pmb \xi, \textbf p)  \,, \\
& = \sum_{ \textbf M \in \mathbb Z^{3n}\!/\{\textbf 0\} } \int_q e^{i \textbf q \cdot (\pmb \xi + L \textbf n )} \int_{k_1} e^{i \textbf k_1 \cdot L \textbf m_1} \cdots \int_{k_{n-1}} e^{i \textbf k_{n-1}\cdot L \textbf m_{n-1}} (-i)^{n} D^{(d)}(p,q, k_1, \cdots, k_n) \,, \\
& =   \sum_{ \textbf M \in \mathbb Z^{3n}\!/\{\textbf 0\} } \int_{q_E} e^{i \textbf q \cdot (\pmb \xi + L \textbf n )} \int_{k_{1,E}} e^{i \textbf k_1 \cdot L \textbf m_1} \cdots \int_{k_{n-1,E}} e^{i \textbf k_{n-1} \cdot L \textbf m_{n-1}}  D^{(d)}_E(p_E,q_E, k_{1,E}, \cdots, k_{n,E}) \,,
\end{align}
where $\textbf M = \{ \textbf n, \textbf m_1, \cdots , \textbf m_{n-1} \}$ and the notation under the sum indicates that the only point omitted is when all three vectors vanish. Introducing $K_E \equiv \{q_E, k_{1,E}, \cdots , k_{n-1,E} \}$ we reach a very compact form for the residue
\begin{equation}
\delta \mathcal M^{(d)}_{L}(\pmb \xi, \textbf p) =   \sum_{ \textbf M \in \mathbb Z^{3n}\!/\{\textbf 0\} } \int_{K_E}  e^{i \textbf q \cdot \pmb \xi + i \textbf K \cdot L \textbf M}   D^{(d)}_E(p_E,K_{E}) \,. \label{eq:master}
\end{equation}
Heuristically, $\textbf M$ parametrizes the space of images that enforce the finite-volume boundary conditions, and the smallest nonzero values (the nearest neighbors) give the dominant finite-volume effects. These can be in the $\textbf n$ direction, corresponding to effects on the Fourier transform from $\textbf q$ to $\pmb \xi$, as well as the $\textbf m_i$ directions, corresponding to finite-volume effects within the diagram.

Returning again to the leading-order diagram, Fig.~\ref{fig:rules_LO}(b), and using the Euclidean form of Eq.~\eqref{eq:master}, we reach 
\begin{align}
\delta \mathcal M^{(\text{LO} )}_{L}(\pmb \xi, \textbf p) & =    g_\varphi^2\sum_{ \textbf n\neq 0 } \int_{q_E}  e^{i \textbf q \cdot (\pmb \xi + i L \textbf n)}   \frac{1}{(p_E+q_E)^2 + m_\varphi^2 } \,.
\end{align}
In Appendix~\ref{App:Igamma}, we review standard tools for writing these integrals in terms of modified Bessel functions. In particular, we find it convenient to define
\begin{equation}
\mathcal{I}_\gamma \big [  \vert  \pmb \xi    \vert     ; m \big ]  \equiv  \int_{k_E}   \frac{e^{ i\textbf{k}\cdot   \pmb \xi}  }{[ k_E^2 + m^2  ]^\gamma} =  \frac{1}{8 \pi^2 \Gamma(\gamma)}
\, \left(\frac{ \vert  \pmb \xi    \vert }{2m}\right)^{\gamma-2}  K_{\gamma-2}\left({ \vert  \pmb \xi    \vert  m}\right) \,
\label{eq:Igamma0},
\end{equation}
implying
\begin{align}
\delta \mathcal M^{(\text{LO} )}_{L}(\pmb \xi, \textbf p) 
&  =  
g_\varphi^2
\sum_{ \textbf n\neq 0}
e^{-i\textbf{p}\cdot  (\pmb \xi+L\textbf n)}  
\mathcal{I}_{1} \big [  |\pmb \xi+L\textbf n|      ; m_\varphi\big ]  \,,
\\
&  =   \frac{ m_{\varphi} g_\varphi^2}{4 \pi^2} 
e^{-i\textbf{p}\cdot  \pmb \xi }
\sum_{ \textbf n\neq 0}  
\frac{K_{1} \big (m_{\varphi} {|\pmb \xi+L\textbf n| } \big)}{|\pmb \xi+L\textbf n|} \,,
\label{eq:MLO_p}
\end{align}
where in the second step we used that $\textbf p = (2 \pi /L) \textbf m$ and therefore that $\exp(- i \textbf p \cdot L \textbf n) = 1$. We comment that, up to a proportionality constant, this is just the finite-volume meson propagator in position space. (See also Ref.~\cite{Cherman:2016vpt}.) 

Keeping only the $\textbf n = - \hat{\pmb \xi}$ term, we find that the dominant finite-volume effect scales as
\begin{equation}\label{eq:m1bphi}
\delta \mathcal{M}^{(\text{LO})}_L ( \pmb \xi ,  \textbf p) =  \frac{ m_{\varphi} g_\varphi^2}{4 \pi^2}  e^{-i\textbf{p}\cdot  \pmb \xi }
\frac{K_{1} \big (m_{\varphi} {| L - \xi | } \big)}{|L -  \xi |}  \longrightarrow  \frac{ m_\varphi^2 g_\varphi^2}{4 \sqrt{2} \pi ^{3/2}} e^{-i\textbf{p}\cdot  \pmb \xi }  \, \frac{ e^{- m_\varphi (L- \xi)}}{ [m_\varphi (L- \xi) ]^{3/2}} \,,
\end{equation}
where the arrow indicates the asymptotic limit.%
\footnote{For fixed $\pmb \xi$ and fixed $\textbf m$ in $\textbf p = (2 \pi /L) \textbf m$, the phase factor $e^{-i\textbf{p}\cdot  \pmb \xi }$ oscillates as $L$ is varied. Here we have in mind estimating a trajectory of fixed $\textbf p$ and $\pmb \xi$ so that the infinite-volume observable is fixed as $L$ varies. We thus do not count the $L$ dependence within the phase factor. }
 The key scaling is given by stripping off the coupling and other prefactors,
\begin{equation}
{\delta \mathcal{M}^{(\text{LO})}_L ( \pmb \xi ,  \textbf p) }\propto    \frac{e^{- m_\varphi (L-  \xi) }}{(L- \xi)^{3/2}}  \,.
\label{eq:asymp_phi}
\end{equation} 
This is the main result of this section and corresponds to the first term on the right-hand side of Eq.~(\ref{eq:main_result}).

Note that the infinite-volume prediction of this diagram can also be read from this expression by replacing $\vert L - \xi \vert$ with $\xi$. This implies, in particular, that the diagram diverges in the limit $\vert \xi \vert \to 0$, as illustrated in Fig.~\ref{fig:periodic}. However, given that we consider a toy EFT that is necessarily written in terms of hadrons, one cannot expect to accurately describe the behavior of physical amplitudes for short distances of the scale $\xi<m_\varphi^{-1}$. We thus require $m_\varphi \xi  \gtrsim 1$, to ensure that the finite- and infinite-volume matrix elements are well described by the EFT.

We close this section by commenting on the diagram shown in Fig.~\ref{fig:rules_LO}(c). As this is only a contact interaction it introduces no finite-volume effects to the matrix element. To understand this in detail requires including a renormalization factor for the product of currents, to accommodate divergences when the two overlap. We have studied this to ensure that no unexpected issues arise.

\section{Beyond leading order}

\label{sec:heavy}

We now turn our attention to the case that the heavy particle, denoted $\chi$, appears in the external state. The leading-order contribution to this matrix element is given by Diagram \ref{fig:rules_LO}(b), with the dotted $\varphi$ propagator replaced by a $\chi$ propagator, and the result is Eq.~\eqref{eq:m1bphi} with the substitutions $m_\varphi \to m_\chi$ and $g_\varphi \to g_\chi$,
\begin{equation}\label{eq:m1bchi}
\delta \mathcal{M}^{(\text{LO})}_L ( \pmb \xi ,  \textbf p) =  \frac{ m_{\chi} g_\chi^2}{4 \pi^2} 
e^{-i\textbf{p}\cdot  \pmb \xi }
\frac{K_{1} \big (m_{\chi} {| L - \xi | } \big)}{|L -  \xi |}  \longrightarrow  
\frac{ m_\chi^2 g_\chi^2}{4 \sqrt{2} \pi ^{3/2}} e^{-i\textbf{p}\cdot  \pmb \xi }
 \, \frac{ e^{- m_\chi (L- \xi)}}{ [m_\chi (L- \xi) ]^{3/2}} \,.
\end{equation}
We take $m_\chi \gg m_\varphi$ and $m_\varphi L \gg 1$, implying that effects of $\mathcal{O}\left(e^{- m_\chi (L- \xi)}\right)$ can be safely ignored. In particular, as we show in this section, the leading finite-volume effects the for the matrix element with a $\chi$ external state are generated by the next-to-leading-order corrections to this result.

\begin{figure}[t]
\centering
\includegraphics[width =1\textwidth]{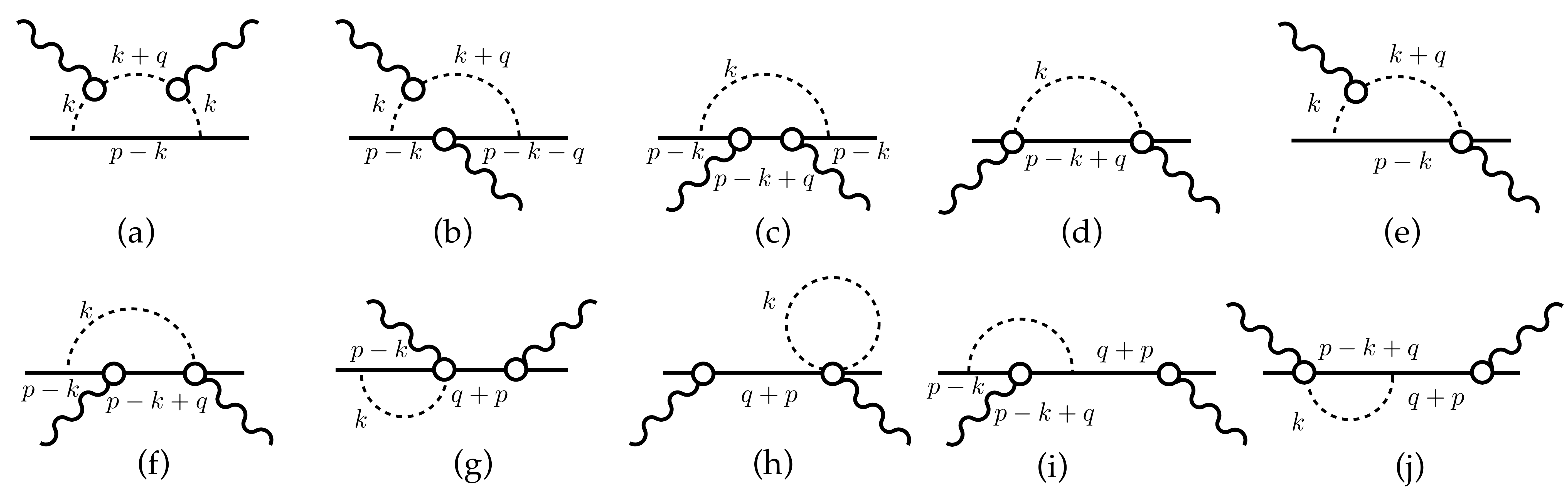}
\caption{Next-to-leading-order contributions to the matrix elements when the external state is the heavy particle. Corrections to the wave function renormalization of the external states are not shown. }\label{fig:chi_NLO} 
\end{figure}

In Fig.~\ref{fig:chi_NLO} we show the one-loop corrections to the matrix element. Here we omit diagrams that give finite-volume corrections to the external states. These give corrections of $\mathcal{O}(e^{-m_\varphi L})$ and, since we are interested in volume effects enhanced by the nonlocality scale $\xi$, can be safely dropped. We first derive general integral expressions for the diagrams in Fig.~\ref{fig:chi_NLO}, restricting attention to the case where the external particle is at rest in the finite volume and highlighting Diagram \ref{fig:chi_NLO}(a) as a specific example. Generally, the integrals that arise in evaluating these diagrams cannot be carried out analytically. To study their asymptotic behavior, in Sec.~\ref{sec:asymptotics} we separate the expressions into analytic parts that dominate the volume scaling together with numerically determined functions that are slowly varying and give only subleading corrections to the scaling.

\bigskip

\paragraph{Diagram~\ref{fig:chi_NLO}(a):}

We illustrate the calculation of the one-loop diagrams in Fig.~\ref{fig:chi_NLO} by outlining the derivation of finite-volume effects for Diagram~\ref{fig:chi_NLO}(a). The calculation of Diagrams~\ref{fig:chi_NLO}(b) to~\ref{fig:chi_NLO}(h) proceeds in a similar fashion but Diagrams~\ref{fig:chi_NLO}(i) and~\ref{fig:chi_NLO}(j) require special treatment, as we discuss below and in Appendix~\ref{sec:diagram_i_j}. From Eq.~(\ref{eq:master}), we identify the finite-volume residue for \ref{fig:chi_NLO}(a) as
 \begin{align} 
\delta \mathcal{M}^{(a)}_L ( \pmb \xi ,  \textbf p)
=
 g^2 g_\varphi^2 \sum_{\{ \textbf{n}, \textbf {m} \} \neq \textbf 0}
\int_{q_E,k_E}
e^{i\textbf{q}\cdot  (\pmb \xi+L\textbf n)}  e^{i L \textbf{k}\cdot  \textbf m}    \frac{1}{[ k_E^2 + m_\varphi^2  ]^2}
\frac{1}{(k_E+q_E)^2 + m_\varphi^2 }    \frac{1}{(p_E-k_E)^2 + m_\chi^2 }   \,,
\end{align}
where the notation below the summation indicates that only the $\textbf n = \textbf m = 0 $ term is omitted from the sum. We separate the $k_E$ and $q_E$ dependence by shifting $q_E \to q_E-k_E$ to reach
 \begin{align} 
\delta \mathcal{M}^{(a)}_L ( \pmb \xi ,  \textbf p)
=
 g^2 g_\varphi^2 \sum_{\{ \textbf{n}, \textbf {m} \} \neq \textbf 0}
\int_{q_E}
 \frac{e^{i\textbf{q}\cdot  (\pmb \xi+L\textbf n)}}{q_E^2 + m_\varphi^2  }    \int_{k_E}   \frac{e^{ i\textbf{k}\cdot  [L(\textbf m - \textbf n) - \pmb \xi]}  }{[ k_E^2 + m_\varphi^2  ]^2[(p_E-k_E)^2 + m_\chi^2 ]}
   \,.
\end{align}

Next, we use a Feynman parameter to reduce the second integral. Starting with the identity 
\begin{equation}
  x (k_E^2 + m_\varphi^2) + (1-x) [ (p_E-k_E)^2 + m_\chi^2 ]  =  ( k_E - (1-x) p_E)^2  + x  m_\varphi^2 + (1-x)  m_\chi^2   + x (1-x) p_E^2  \,,
\end{equation}
we shift $k_E \to k_E + (1 - x) p_E$ to reach
 \begin{align} 
\delta \mathcal{M}^{(a)}_L ( \pmb \xi ,  \textbf p)
=
2 g^2 g_\varphi^2 \int_0^1 \mathrm{d}x x \sum_{\{ \textbf{n}, \textbf {m} \} \neq \textbf 0}
e^{ i (1-x)\textbf{p}\cdot  [L(\textbf m - \textbf n) - \pmb \xi]}
\int_{q_E}
 \frac{e^{i\textbf{q}\cdot  (\pmb \xi+L\textbf n)}}{q_E^2 + m_\varphi^2  }    \int_{k_E}   \frac{e^{ i\textbf{k}\cdot  [L(\textbf m - \textbf n) - \pmb \xi]}  }{[ k_E^2 + M(x)^2  ]^3} \,,
\end{align}
where
\begin{equation}
M(x)^2 \equiv  x  m_\varphi^2 + (1-x)  m_\chi^2   + x (1-x) p_E^2 =  x  m_\varphi^2 + (1-x)^2  m_\chi^2    \,.
\label{M_x}
\end{equation}
In the second step, we have set the Euclidean external momentum on-shell, $p_E^2 = - m_\chi^2$.

At this stage, we have written the loop in terms of products of two integrals of the kind given in Eq.~(\ref{eq:Igamma0}). Substituting the definition of $\mathcal I_\gamma$ then gives 
 \begin{align} 
\delta \mathcal{M}^{(a)}_L ( \pmb \xi ,  \textbf p)
=
 2 g^2 g_\varphi^2 \int_0^1 \mathrm{d}x x \sum_{\{ \textbf{n}, \textbf {m} \} \neq \textbf 0}  e^{ i (1-x)\textbf{p}\cdot  [L \textbf m  - \pmb \xi]}
  \, \mathcal{I}_1 \big [ \vert  L\textbf n  - \pmb \xi  \vert ;m_\varphi \big ]  \, \mathcal{I}_3 \big [  \vert    L \textbf m - \pmb  \xi     \vert     ; M(x) \big ]   
   \,,
   \label{eq:3Aalmost}
\end{align}
where we have shifted the summed integer vectors. Taking the external state to be at rest in the finite volume, i.e.~setting $\textbf p= \textbf 0$, then gives 
\begin{equation}
\delta \mathcal{M}^{(a)}_L ( \pmb \xi ,  \textbf 0)
=
 2 g^2 g_\varphi^2 \sum_{\{ \textbf{n}, \textbf {m} \} \neq \textbf 0} 
  \, \mathcal{I}_1 \big [ \vert L\textbf n -  \pmb \xi  \vert ;m_\varphi \big ]  \, \left[\int_0^1 \mathrm{d}x x\, \mathcal{I}_3 \big [  \vert L \textbf m -  \pmb \xi   \vert     ; M(x) \big ]   \right]
   \,.
   \label{eq:3Aalmost2}
\end{equation}

 \paragraph{Diagrams~\ref{fig:chi_NLO}(b) to \ref{fig:chi_NLO}(h)\label{ssec:btoh}:}

This set of diagrams is amenable to the same approach as Diagram~\ref{fig:chi_NLO}(a). In Appendix~\ref{App:loopsIgamma}, we present a simple generalization of the technique presented above for Diagram~\ref{fig:chi_NLO}(a) that allows for a rapid derivation of the finite-volume effects for these diagrams. The results for $\textbf p = \textbf 0$ are  
 \begin{align} 
 \delta \mathcal{M}^{(b)}_L ( \pmb \xi ,  \textbf 0)
& =   g^2 g_\varphi g_\chi 
\sum_{\{ \textbf{n}, \textbf {m} \} \neq \textbf 0}
 \left[\int_0^1 \mathrm{d}x     \, \mathcal{I}_2 \big [ \vert  L\textbf n - \pmb \xi  \vert ;M(x) \big ] 
\right] \,
\left[ \int_0^1 \mathrm{d}y  \,  \mathcal{I}_2 \big [  \vert  L   \textbf m   - \pmb \xi  \vert     ; M(y) \big ]
\right]
\,,\label{eq:3Balmost} \\
  \delta \mathcal{M}^{(c)}_L ( \pmb \xi ,  \textbf 0)
& =  2 g^2 g_\chi^2    \sum_{\{ \textbf{n}, \textbf {m} \} \neq \textbf 0}
  \, \mathcal{I}_1 \big [ \vert  L\textbf n -  \pmb \xi  \vert ;m_\chi \big ]  \, 
  \left[ \int_0^1 \mathrm{d}x\,(1-x)\,\mathcal{I}_3 \big [  \vert L  \textbf m -  \pmb \xi    \vert     ; M(x) \big ] \right] \,,\label{eq:3Calmost} \\
\delta \mathcal{M}^{(d)}_L ( \pmb \xi ,  \textbf 0)
& =  g_{\chi\varphi}^2  \sum_{\{ \textbf{n}, \textbf {m} \} \neq \textbf 0}
  \, \mathcal{I}_1 \big [ \vert L\textbf n - \pmb \xi  \vert ;m_\chi \big ]  \, \mathcal{I}_1 \big [  \vert L  \textbf m -  \pmb \xi    \vert     ; m_\varphi \big ]\,,\label{eq:3Dalmost} \\
\delta \mathcal{M}^{(e)}_L ( \pmb \xi ,  \textbf 0)
& =  g g_\varphi g_{\chi\varphi} \sum_{\{ \textbf{n}, \textbf {m} \} \neq \textbf 0}
  \, \mathcal{I}_1 \big [ \vert L\textbf n -  \pmb \xi  \vert ;m_\varphi \big ]  \, \left[  \int_0^1 \mathrm{d}x\,     \mathcal{I}_2 \big [  \vert  L  \textbf m -  \pmb \xi    \vert     ; M(x) \big ]\right] \,,\label{eq:3EFalmost} \\
\delta \mathcal{M}^{(f)}_L ( \pmb \xi ,  \textbf 0)
& =  g g_\chi g_{\chi\varphi}     \sum_{\{ \textbf{n}, \textbf {m} \} \neq \textbf 0}
  \, \mathcal{I}_1 \big [ \vert L\textbf n -  \pmb \xi  \vert ;m_\chi \big ]  \left[ \int_0^1 \mathrm{d}x \, \mathcal{I}_2 \big [  \vert L  \textbf m -   \pmb \xi     \vert     ; M(x) \big ]\right] \,,\label{eq:3Falmost} \\
\delta \mathcal{M}^{(g)}_L ( \pmb \xi ,  \textbf 0)
& =   g g_{\chi\varphi}g_{\chi}    \sum_{\{ \textbf{n}, \textbf {m} \} \neq \textbf 0}
  \, \mathcal{I}_1 \big [ \vert   L\textbf n - \pmb \xi  \vert ;m_\chi \big ]  \,
  \left[   \int_0^1 \mathrm{d}x\, \mathcal{I}_2 \big [  \vert  L \textbf m \vert     ; M(x) \big ]\right] \,,\label{eq:3Galmost} \\
\delta \mathcal{M}^{(h)}_L ( \pmb \xi ,  \textbf 0)
& = \frac12 g_{\chi} g_{\chi\varphi\varphi} \sum_{\{ \textbf{n}, \textbf {m} \} \neq \textbf 0}
  \, \mathcal{I}_1 \big [ \vert L\textbf n -  \pmb \xi  \vert ;m_\chi \big ]  \, \mathcal{I}_1 \big [  \vert  L \textbf m \vert     ; m_\varphi \big ]\,.\label{eq:3Halmost}
\end{align}

The key feature for these diagrams is that one can factorize the dependence on the current momentum, $q$, from that on the internal loop momentum, $k$. In all cases, this results in two modified Bessel functions, corresponding to the two momenta after an appropriate shift has been performed. Note that the sum of the indices on the two Bessel functions always corresponds to the number of internal propagators.

\paragraph{Diagrams~\ref{fig:chi_NLO}(i) and (j)\label{ssec:ij}}

These two diagrams cannot be factorized into two separate momentum integrals and must be studied using a different approach. 
In Appendix~\ref{sec:diagram_i_j}, we evaluate these diagrams and place upper bounds on their values. We demonstrate that the finite-volume artifacts associated with these are smaller than those for \ref{fig:chi_NLO}(a) to~\ref{fig:chi_NLO}(h). As we are only interested in the dominant finite-volume effects, we ignore the contributions from Figs.~\ref{fig:chi_NLO}(i) and~\ref{fig:chi_NLO}(j) from here on.

\section{Asymptotic behavior\label{sec:asymptotics}}

In this section we study the asymptotic behavior of Eqs.~\eqref{eq:3Aalmost2} through \eqref{eq:3Halmost}. As mentioned above, we assume that $m_\chi\gg m_\varphi$ and ignore corrections that decrease with the volume as $e^{-m_\chi L}$ or more rapidly. As mentioned above, the matrix element must be periodic with periodicity $L$. Thus, as $\xi$ approaches $L$, the finite-volume effects become arbitrarily large [see also Fig.~\ref{fig:periodic} above]. Here we are not directly interested in this regime of extreme volume effects but rather in the region of $\xi = c L$ with $c \ll 1$. 
This motivates us to take the asymptotic forms of the Bessel functions, i.e.~to take the arguments $\vert L \textbf n - \pmb \xi \vert$ as large. 
 
Combining the asymptotic form of the Bessel functions with the definition of $\mathcal I_{\gamma},$ Eq.~(\ref{eq:Igamma0}), we find 
\begin{align}
\mathcal{I}_\gamma \big [  \vert \pmb z    \vert     ; m \big ] 
&=  \frac{1}{8 \pi^{3/2} \Gamma(\gamma)}
{\frac{(2m)^{3/2-\gamma}}{  \vert \pmb z    \vert^{5/2-\gamma}}} e^{- m \vert \pmb z    \vert  } \bigg [1 + \mathcal O \bigg ( \! \frac{1}{m \vert \pmb z \vert} \!  \bigg ) \bigg ]\,.
\end{align}
Given this exponential suppression, terms with $\textbf n$ chosen to minimize $\vert L \textbf n - \pmb \xi \vert$ will dominate the sum. In addition, terms scaling as $e^{- m_{\chi} \vert L \textbf n - \pmb \xi \vert}$, i.e.~with the mass of the heavier particle, will be highly suppressed and we drop such contributions throughout.

In Eqs.~\eqref{eq:3Aalmost2} through \eqref{eq:3Halmost}, only $\gamma=1,2,3$ appear. We thus give their explicit forms for convenience,
\begin{align}
\mathcal{I}_1 \big [  \vert \pmb z    \vert     ; m \big ] &\sim  \frac{1}{8 \pi^{3/2} }
{\frac{(2m)^{1/2}}{ \vert \pmb z    \vert^{3/2}}} e^{- \vert \pmb z    \vert  m} \,,
\\
\mathcal{I}_2 \big [  \vert \pmb z    \vert     ; m \big ] &\sim \frac{1}{8 \pi^{3/2}}
{\frac{e^{- \vert \pmb z    \vert  m}}{ \sqrt{2m \vert \pmb z    \vert}}}  \,,
\\
\mathcal{I}_3 \big [  \vert \pmb z    \vert     ; m \big ] &\sim  \frac{1}{16 \pi^{3/2} }
{\frac{ \vert \pmb z    \vert^{1/2}}{ (2m)^{3/2}}} e^{-\vert \pmb z    \vert  m} \,,
\label{eq:asymp}
\end{align}
where we use $\sim$ to indicate that the two sides agree up to terms suppressed by additional powers of $1/(m \vert  \pmb z \vert)$.

The asymptotic forms of the one-loop diagrams can be determined using a similar approach to that for the leading-order diagram, Eq.~(\ref{eq:m1bchi}). We identify the dominant terms in the sums over $\textbf n$ and $\textbf m$ assuming $\xi = c L$ with $c \ll 1$. The only additional subtlety is that the integrals over Feynman parameters are found to be numerically dominated by $M(x) \sim m_{\varphi}$. Factoring out this dependence, we reach the following,
 \begin{align} 
 \delta \mathcal{M}^{(a)}_L ( \pmb \xi ,  \textbf 0)
& \sim
  \frac{g^2 g_\varphi^2 }{128 \pi^{3} m_{\varphi} }   \bigg [ {\frac{\xi^{1/2}}{   (L - \xi)^{3/2}}}   H_{x,3/2}(\xi  )   +   {\frac{(L - \xi)^{1/2}}{     \xi^{3/2}}}   H_{x,3/2}(L - \xi  )       \bigg ] e^{-  m_{\varphi}  L  }
   \,, \label{eq:3A}\\
 \delta \mathcal{M}^{(b)}_L ( \pmb \xi ,  \textbf 0)
& \sim   \frac{g^2 g_\varphi g_\chi }{64 \pi^3 m_{\varphi}} \bigg [ \frac{1}{\xi^{1/2} (L - \xi)^{1/2} }      H_{1,1/2}(\xi  )     H_{1,1/2}(L - \xi  )  \bigg ]  e^{- m_{\varphi} L}
\,,\label{eq:3B} \\
  \delta \mathcal{M}^{(c)}_L ( \pmb \xi ,  \textbf 0)
& =  \frac{g^2 g_\chi^2 }{128 \pi^3} \frac{m_{\chi}^{1/2}}{m_{\varphi}^{3/2} }  \bigg [ \frac{(L - \xi)^{1/2}}{\xi^{3/2}}    H_{1-x,3/2}(L - \xi) \bigg ] e^{- \xi (m_{\chi} - m_{\varphi} )}
 e^{- m_{\varphi}  L }    \, 
  \,,\label{eq:3C} \\
\delta \mathcal{M}^{(d)}_L ( \pmb \xi ,  \textbf 0)
& = \frac{ g_{\chi\varphi}^2 m_\chi^{1/2} m_{\varphi}^{1/2}}{32 \pi^3}  \bigg [   \frac{1}{\xi^{3/2} (L - \xi)^{3/2}} \bigg ]
 e^{- \xi (m_{\chi} - m_{\varphi} )}
 e^{- m_{\varphi}  L }\,,\label{eq:3D} \\
\delta \mathcal{M}^{(e)}_L ( \pmb \xi ,  \textbf 0)
& =  \frac{g g_\varphi g_{\chi\varphi} }{64 \pi^3}  \bigg [  \frac{1}{\xi^{1/2} (L - \xi)^{3/2}}  H_{1,1/2}(\xi)+  \frac{1}{\xi^{3/2} (L - \xi)^{1/2}}    H_{1,1/2}(L - \xi)     \bigg ]
  \,    e^{- m_{\varphi} L} \,,\label{eq:3EF} \\
\delta \mathcal{M}^{(f)}_L ( \pmb \xi ,  \textbf 0)
& =  \frac{g g_\chi g_{\chi\varphi} m_{\chi}^{1/2} }{64 \pi^3 m_{\varphi}^{1/2}} \bigg [  \frac{1}{\xi^{3/2} (L - \xi)^{1/2}}  H_{1,1/2}(L-\xi) \bigg ] e^{- \xi (m_{\chi} - m_{\varphi} )}
 e^{- m_{\varphi}  L } \,,\label{eq:3F} \\
\delta \mathcal{M}^{(g)}_L ( \pmb \xi ,  \textbf 0)
& =   \frac{g g_{\chi\varphi}g_{\chi} m_{\chi}^{1/2} }{64 \pi^3 m_{\varphi}^{1/2}}  \bigg [ \frac{1}{\xi^{3/2} L^{1/2}}   H_{1,1/2}(L) \bigg ] e^{- \xi m_{\chi} }
 e^{- m_{\varphi}  L } \,,\label{eq:3G} \\
\delta \mathcal{M}^{(h)}_L ( \pmb \xi ,  \textbf 0)
& =  \frac{g_{\chi} g_{\chi\varphi\varphi} m^{1/2}_{\varphi}  m^{1/2}_{\chi}  }{64 \pi^3} \bigg [ \frac{1}{\xi^{3/2} L^{3/2}}  \bigg ]  e^{- m_{\chi} \xi} e^{- m_{\varphi} L}  \,,\label{eq:3H}
\end{align}
where
\begin{equation}
H_{f(x),\alpha}(\xi ) = \int_0^1 \mathrm{d}x f(x)\, \frac{m_{\varphi}^{\alpha}}{M(x)^{\alpha}}  e^{-  \xi  (M(x) - m_{\varphi} )}  \,.
\end{equation}

\begin{figure}[t]
\begin{center}
\includegraphics[width =0.7\textwidth]{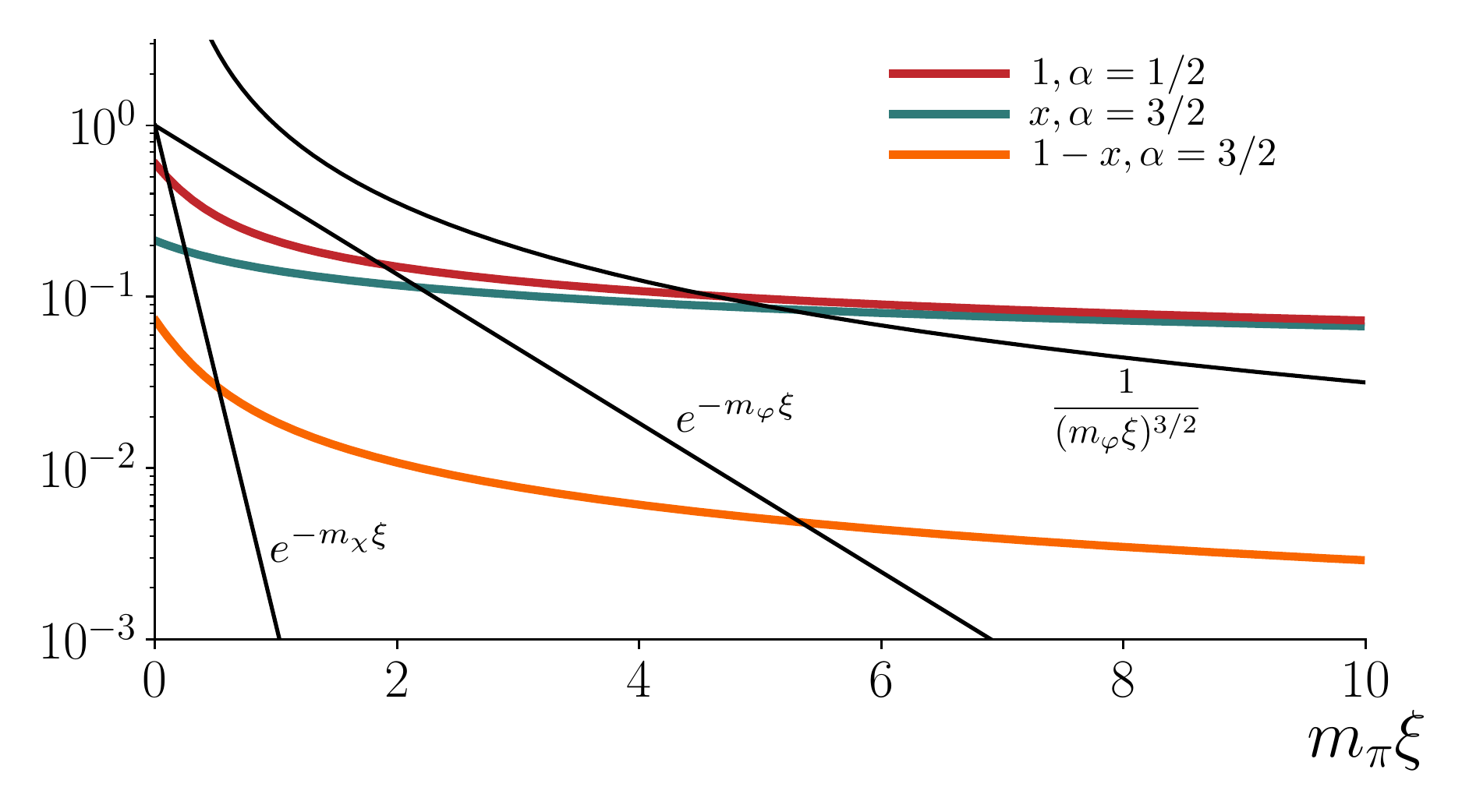}  
\vspace{-10pt}
\caption{Plot of the function $H_{f(x),\alpha}(\xi)$ vs $m_{\varphi} \xi$ for $m_{\chi} /m_{\varphi}$ set to the nucleon-pion mass ratio. The figure shows that the $H$ functions are slowly varying compared to the corresponding exponentials and powers of $\xi$ appearing in Eqs.~\eqref{eq:3A} to \eqref{eq:3H}. Thus the leading scaling is given by treating these functions as constant. \label{fig:JFig}}  
\end{center}
\end{figure}

As we show in Fig.~\ref{fig:JFig}, $H_{f(x),\alpha}(\xi)$ is a slowly varying function of its argument. Thus, the leading scaling can be read from the given expressions. We deduce that, in all cases, the finite-volume effects are suppressed by a factor of $e^{- m_\varphi L}$. In particular, the dominant finite-volume contributions come from diagrams (a), (b) and (e) with the leading effect for small $\xi$ driven by the $(L - \xi)^{1/2}/\xi^{3/2}$ factor appearing in diagram (a).

\section{Summary\label{sec:summary}}

We have presented the first steps toward understanding the finite-volume artifacts that arise in matrix elements of spatially nonlocal operators. These operators are relevant for a wide variety of observables being studied using lattice QCD, including parton distribution functions, double-beta-decay and Compton-scattering matrix elements, and long-range weak matrix elements. In particular, matrix elements of products of spatially separated currents represent one approach to determining hadron structure directly from lattice QCD \cite{Ma:2017pxb, Braun:2007wv,Bali:2017gfr}.

We considered a toy model involving two scalar particles, one analogous to the pion in QCD and one analogous to the nucleon or a heavy meson, and determined the finite-volume matrix elements of these states with two spatially separated scalar currents at one loop in perturbation theory. As expected, we found that these matrix elements are contaminated by larger finite-volume artifacts than is the case for matrix elements of local operators. The results are summarized in Eq.~\eqref{eq:main_result}. There are two terms that potentially dominate the finite-volume artifacts, one scaling with the mass of the external state and the other with the mass of the lightest degree of freedom. When these two coincide, the dominant finite-volume correction scales as $e^{- m_\pi (L - \xi)}$, provided $\xi$ is a non-negligible fraction of $L$. By contrast, if the external state is significantly heavier than the lightest particle, as in the case of a nucleon or heavy meson, then the leading finite-volume artifacts scale as $e^{- m_\pi L}$. In both cases, these exponential factors have polynomial prefactors, including terms scaling as $L^m/|L - \xi|^n$ that could have a significant impact on the size of finite-volume corrections.

Future extensions of this work include implementing the approach for specific QCD matrix elements using $\chi$PT, including flavor-changing currents, and more carefully studying the dependence on the external state momenta (especially at next-to-leading order).  
%
A more involved extension would be to apply the approach to operators involving Wilson lines, such as those relevant for determining quasi- and pseudo-PDFs. This requires a $\chi$PT-based representation of such operators~\cite{Chen:2001nb, Kivel:2002ia, Kivel:2007jj,Diehl:2005rn}, potentially built around the methods of Ref.~\cite{Green:2017xeu}. Finally, here we have only considered periodic boundary conditions. Previous works~\cite{Briceno:2013hya, Cherman:2016vpt, Korber:2015rce} have shown that particular choices of twisted boundary conditions~\cite{Bedaque:2004kc, Byers:1961zz} can be used to reduce the size of exponentially suppressed finite-volume artifacts. This may also prove useful in reducing volume corrections for matrix elements of spatially nonlocal operators. 

\section*{Acknowledgments}

R.~A.~B.~and J.~V.~G.~acknowledge support from U.S. Department
of Energy Contract No. DE-AC05-06OR23177, under which Jefferson Science Associates,
LLC, manages and operates Jefferson Lab. J.~V.~G.~is supported in part by the U.S.~Department of Energy through DOE Contract No. DE-SC0008791 and also
through the JSA Graduate Fellowship Program.
C.~J.~M.~is supported in part
by the U.S.~Department of Energy through Grant No. DE-FG02-00ER41132. The authors would like to thank Robert Edwards for persistently asking questions that inspired this project. The authors would also like to thank Alberto Accardi, Martin Savage, Joseph Karpie, and David Richards for helpful conversations during the course of this work. 
\appendix

\section{Integrals in terms of modified Bessel functions}
\label{App:Integrals_Bessel}

In Sec.~\ref{sec:heavy} we have shown that complicated diagrams can be written as integrals of products of modified Bessel functions. Although these integrals are well documented and derived in the literature (see, for example, Ref.~\cite{Bijnens:2013doa}), we review the derivation of the necessary functions in this Appendix. 


\subsection{Derivation of Eq.~(\ref{eq:Igamma0}) 
\label{App:Igamma}}

We begin by deriving Eq.~(\ref{eq:Igamma0}), the identity relating the function $\mathcal I_{\gamma}$,
\begin{equation}
\label{eq:Igammadef}
\mathcal{I}_\gamma \big [  \vert  \pmb \xi    \vert     ; m \big ]  \equiv  \int_{k_E}   \frac{e^{ i\textbf{k}\cdot   \pmb \xi}  }{[ k_E^2 + m^2  ]^\gamma} \,, \end{equation}
to the modified Bessel function, $K_{\gamma-2}$.

Beginning with the definition of the $\Gamma$ function
 \begin{align}
 \frac{1}{Q^{\gamma}}=\frac{1}{\Gamma(\gamma)}\int^\infty_0 d\alpha \, e^{-\alpha Q}\alpha^{\gamma-1} \,,
 \label{eq:Schwinger_param}
 \end{align}
we observe
 \begin{align}
\mathcal{I}_\gamma[\xi;m]=
  \int_{k_E} \,e^{i \textbf k \cdot  \pmb \xi}
\frac{1}{\Gamma(\gamma)}\int^\infty_0 d\alpha \, e^{-\alpha (k^2+m^2)}\alpha^{\gamma-1} \,.
 \end{align}
 
Next, we complete the square in the four-vector, $k_E$, to write
\begin{align}
-\alpha k^2_E+i \textbf k\cdot  \pmb \xi
=
-\alpha \left(\left(k_E^\mu-i\frac{  \xi^\mu}{2\alpha}\right)^2 +\left(\frac{ \xi^\mu}{2\alpha}\right)^2\right) \,,
\end{align}
where $  \xi^\mu=(0, \pmb \xi)$. Performing the integral over $k_E$ then gives
 \begin{align}
\mathcal{I}_\gamma[\xi;m]
&=
\frac{1}{\Gamma(\gamma) (4\pi)^2}
\,
 \,
\int^\infty_0 d\alpha \, e^{-\alpha m^2-\frac{\xi^2}{4\alpha}}\alpha^{\gamma-3} \,,
 \end{align}
 where we set $\xi=|\pmb \xi|$ from here on.
 
 Finally, we perform the variable substitution $ \alpha =  \xi e^{\theta} / (2m)$ to reach
\begin{align}
\mathcal{I}_\gamma[\xi;m] 
&=
\frac{1}{\Gamma(\gamma)(4\pi)^2}
\,
\left(\frac{\xi}{2m}\right)^{\gamma-2}
\int^\infty_{-\infty} d\theta \, e^{-{\xi m}\frac{ e^{\theta}+ e^{-\theta}}{2}}e^{(\gamma-2)\theta} \,,
\\ 
&=
\frac{1}{\Gamma(\gamma)(4\pi)^2}
\,
\left(\frac{\xi}{2m}\right)^{\gamma-2}
\int^\infty_{-\infty} d\theta \, e^{-{\xi m}\cosh \theta} \cosh[(\gamma-2)\theta]  \,,
\\ 
&=
\frac{1}{8 \pi^2 \Gamma(\gamma) }
\,
\left(\frac{\xi}{2m}\right)^{\gamma-2}
 \,K_{\gamma-2}\left({\xi m}\right) \,.
\label{eq:Igamma}
\end{align}
Here we have used the fact the $\cosh$ and $\sinh$ are symmetric and antisymmetric respectively, and have introduced the modified Bessel function, $K_{\gamma-2}(z)$. Note that $K_{a}(z) = K_{-a}(z)$.

\subsection{Loops in terms of $\mathcal{I}_\gamma$
\label{App:loopsIgamma}}
Figures~\ref{fig:chi_NLO}(a)-(h) can be written as integrals over products of $\mathcal I_\gamma$ defined in Eq.~(\ref{eq:Igammadef}). Here we show our general method for doing this for all integrals of the form  
\begin{align}
J_{\gamma\gamma'}\equiv\int_{k_E}   \frac{e^{ i\textbf{k}\cdot   \pmb \xi}  }{[ k_E^2 + m_\varphi^2  ]^\gamma} \frac{1}{[ (p_E-k_E)^2 + m_\chi^2  ]^{\gamma'}} \,.
\end{align}
First one inserts a Feynman parameter integral to combine the denominators
\begin{align}
J_{\gamma\gamma'}
&=\frac{\Gamma(\gamma+\gamma')}{\Gamma(\gamma)\Gamma(\gamma')}
\int_0^1 dx \, x^{\gamma-1}\, (1-x)^{\gamma'-1}
e^{ i(1-x)\textbf{p} \cdot   \pmb \xi}
 \int_{k_E}   \frac{e^{ i\textbf{k}\cdot   \pmb \xi}  }{[ k_E^2 + M(x)^2]^{\gamma+\gamma'}}  \,,
\end{align}
where we performed the variable transformation $k_E\to k_E + p_E(1-x)$, used the on-shell condition for the external states, $p_E^2 = - m_\chi^2$, and also substituted $M(x)^2 =  xm_\varphi^2 +m_\chi^2(1-x)^2$.

Using the functions defined in Eq.~(\ref{eq:Igamma}), we arrive at
\begin{align}
J_{\gamma\gamma'}
&=
\frac{\Gamma(\gamma+\gamma')}{\Gamma(\gamma)\Gamma(\gamma')}
\int_0^1 dx \, x^{\gamma-1}\, (1-x)^{\gamma'-1}
e^{ i(1-x)\textbf{p} \cdot   \pmb \xi}
\, \mathcal{I}_{\gamma+\gamma'}\left[\xi;   M(x) \right]
\label{eq:Jggp}
.
\end{align}
From this, it is straightforward to arrive at the expressions given for Figs.~\ref{fig:chi_NLO}(b)-3(h) in Eqs.~(\ref{eq:3Balmost})-(\ref{eq:3Halmost}).

\subsection{Detailed calculation of diagrams (i) and (j)}\label{sec:diagram_i_j}

\paragraph{Diagram~\ref{fig:chi_NLO}(j):}

The contribution of Diagram~\ref{fig:chi_NLO}(j) is given by
\begin{align} 
\delta \mathcal{M}^{(j)}_L ( \pmb \xi ,  \textbf p)
&=
g g_\chi g_{\chi \varphi} \sum_{\{ \textbf{n}, \textbf {m} \} \neq \textbf 0}
\int_{q_E,k_E}
e^{i\textbf{q}\cdot  (\pmb \xi+L\textbf n)}  e^{i L \textbf{k}\cdot  \textbf m}    \frac{1}{ k_E^2 + m_\varphi^2  }
\frac{1}{( p_E - k_E + q_E)^2 + m_\chi^2 }    \frac{1}{(q_E + p_E)^2 + m_\chi^2 }   \, ,
\\
&=
g g_\chi g_{\chi \varphi} \sum_{\{ \textbf{n}, \textbf {m} \} \neq \textbf 0}
\int_{q_E}  e^{i({\bf q-p})\cdot  (\pmb \xi+L\textbf n)}  \frac{1}{q_E^2 + m_\chi^2 } 
\int_{k_E}
e^{i L \textbf{k}\cdot  \textbf m} \frac{1}{ k_E^2 + m_\varphi^2} \frac{1}{( q_E - k_E)^2 + m_\chi^2 }    
\,, 
\end{align}
where in the second step we performed the variable transformation $q_E \to q_E - p_E$. 

We can rewrite the integral over $k_E$ using a Feynman parameter, 
\begin{equation}
x (k_E^2 + m_\varphi^2) + (1-x) [ (k_E-q_E)^2 + m_\chi^2 ]  =  ( k_E - (1-x) q_E)^2 + x (1-x) ( q_E^2 + m_\chi^2 ) +  M(x)^2 \,,
\end{equation}
where $ M(x)^2 \equiv  x  m_\varphi^2 + (1-x)^2  m_\chi^2$, as in the main text. After shifting $k_E \to k_E + (1 - x) q_E$, introducing $\ximn^{(j)} = \xi + L \textbf n + L (1-x)\textbf m
$, and setting ${\bf p}=0$ we arrive at
\begin{align} 
\delta \mathcal{M}^{(j)}_L ( \pmb \xi ,  \textbf 0)
&= 
g g_\chi g_{\chi \varphi} \int_{0}^{1} dx
  \sum_{\{ \textbf{n}, \textbf {m} \} \neq \textbf 0}
\int_{q_E}  e^{i {\bf q} \cdot \ximn^{(j)}}   \frac{1}{q_E^2 + m_\chi^2 } 
\int_{k_E}
\frac{e^{i L \textbf{k}\cdot  \textbf m}}{[k_E^2 + x (1-x) (q_E^2 + m_\chi^2) +  M(x)^2]^2}    
\,.
\end{align}
We then use the Schwinger parametrization, Eq.~\eqref{eq:Schwinger_param}, to reach
\begin{align} 
\delta \mathcal{M}^{(j)}_L ( \pmb \xi ,  \textbf 0)
&=
g g_\chi g_{\chi \varphi} \int_{0}^{1} dx \sum_{\{ \textbf{n}, \textbf {m} \} \neq \textbf 0}
\int_{q_E, k_E}  e^{i {\bf q} \cdot \ximn^{(j)}} 
e^{i L \textbf{k}\cdot  \textbf m}
\int_{0}^{\infty} d\alpha \, 
\int_{0}^{\infty} d\beta \, \beta \,
e^{- (\alpha + \beta  x(1-x) ) [ q_E^2 + m_\chi^2 ]}
e^{-\beta [ k_E^2 + M(x)^2 ]}
\label{eq:delta_ML_j} \,.
\end{align}

At this stage, if we perform the change of variables $\alpha = \zeta - z (1-z) \beta$, then we almost reach the integrated product of two of the $\mathcal I$ functions discussed in Appendix~\ref{App:Igamma}. The only difference is that the lower limit on the $\zeta$ integral differs from zero. But since the integrand over $\zeta$ is always positive and $x(1-x)\beta \geq 0$, we can easily impose an upper limit for this contribution:
\begin{align} 
\delta \mathcal{M}^{(j)}_L ( \pmb \xi ,  \textbf 0)
& \leq
g g_\chi g_{\chi \varphi} \int_{0}^{1} dx
\sum_{\{ \textbf{n}, \textbf {m} \} \neq \textbf 0}
\bigg[\int_{q_E} e^{i \textbf q \cdot \ximn^{(j)}}
\int_{0}^{\infty} d\zeta
e^{-\zeta(q_E^2 + m_\chi^2)} \bigg]
\bigg [\int_{k_E}
e^{i L \textbf{k}\cdot  \textbf m}
\int_{0}^{\infty} d\beta 
\beta \,
e^{-\beta (k_E^2 + M(x)^2)} \bigg] \,, \\
& \leq
g g_\chi g_{\chi \varphi} \int_{0}^{1} dx
\sum_{\{ \textbf{n}, \textbf {m} \} \neq \textbf 0}
\mathcal{I}_1 \big [  \vert \ximn^{(j)} (x)   \vert ; m_\chi \big ] 
\mathcal{I}_2 \big [  \vert L \textbf m    \vert ; M(x) \big ] \,.
\end{align}	

We deduce that the $\xi$ dependence only appears in the $m_\chi$-integral, and thus any enhancement due to the nonlocality of the operator is suppressed in this diagram by the heavier particle mass.

\bigskip

\paragraph{Diagram~\ref{fig:chi_NLO}(i):}

The contribution of this diagram is
\begin{align} 
\delta \mathcal{M}^{(i)}_L ( \pmb \xi ,  \textbf p)
&=
g^2 g_\chi^2 \sum_{\{ \textbf{n}, \textbf {m} \} \neq \textbf 0}
\int_{q_E}
e^{i(\textbf{q}-\textbf {p})\cdot  (\pmb \xi+L\textbf n)}  \frac{1}{q_E^2 + m_\chi^2}
\int_{k_E}
e^{i L \textbf{k}\cdot  \textbf m}    \frac{1}{ k_E^2 + m_\varphi^2}  \frac{1}{(p_E-k_E)^2 + m_\chi^2} \frac{1}{(q_E-k_E)^2 + m_\chi^2} \,,
\end{align}
where we have already performed the shift $q_E \to q_E - p_E$. Introducing two Feynman parameters, labeled $x$ and $z$, allows us to combine the three $k_E$-dependent denominators
\begin{multline}
k_E^2 + m_\varphi^2 + x\Big[\big((p_E-k_E)^2 + m_\chi^2 \big)-( k_E^2 + m_\varphi^2 )\Big] + z\Big[\big((q_E-k_E)^2 + m_\chi^2 \big) - (k_E^2 + m_\varphi^2)\Big] \\  = 
\big[k_E-(xp_E+zq_E)\big]^2 - 2 x z p_E \cdot q_E  + z(1-z) (q_E^2 +   m_\chi^2)  + \Delta(x,z)^2 \,,
\end{multline}
where
\begin{align}
\Delta(x,z)^2
& =
x(1-x)p_E^2 - z(1-z)   m_\chi^2 +  m_\varphi^2 + (x+z)(m_\chi^2 - m_\varphi^2)  \,.
\end{align}

Shifting $k_E \to k_E + (x p_E + z q_E )$ and setting ${\bf p = 0}$, we arrive at
\begin{multline}
\delta \mathcal{M}^{(i)}_L ( \pmb \xi ,  \textbf 0)
= 
2 g^2 g_\chi^2 \sum_{\{ \textbf{n}, \textbf {m} \} \neq \textbf 0}
\int_{0}^{1} dx 
\int_{0}^{1-x} dz \,
\int_{q_E}
e^{i \textbf{q}\cdot \ximn^{(i)}} \frac{1}{q_E^2 + m_\chi^2}
 \\ \times \int_{k_E}
e^{i L \textbf{k}\cdot  \textbf m}    \frac{1}{ [k_E^2 - 2 x z p_E \cdot q_E  + z(1-z) (q_E^2 + m_\chi^2)   + \Delta(x,z)^2]^3}  \,,
\end{multline}
where
\begin{align}
\ximn^{(i)} = \xi+L\textbf n + zL \textbf m \,.
\label{eq:ximn_i}
\end{align}

Proceeding as above we now introduce two Schwinger parameters for the denominators to reach \begin{multline}
\delta \mathcal{M}^{(i)}_L ( \pmb \xi ,  \textbf 0)
= 
 g^2 g_\chi^2 \sum_{\{ \textbf{n}, \textbf {m} \} \neq \textbf 0}
\int_{0}^{1} dx 
\int_{0}^{1-x} dz \,
  \int_{q_E,k_E}
e^{i \textbf{q}\cdot  \ximn^{(i)}}
e^{i L \textbf{k}\cdot  \textbf m}  
\\ \times
\int_{0}^{\infty} d\alpha \,
d\beta \,  \beta^2  \,
e^{ i (2 \beta x z) m_\chi q_E^0} \ e^{-\beta(k_E^2+ {\Delta}^2)}  \ 
e^{- (\alpha + z(1-z) \beta) (q_E^2 + m_\chi^2 )}  \,,
\end{multline}
where we have substituted $p_E \cdot q_E = i m_\chi q_E^0$ and also set $p_E^2$ within $\Delta(x,z)$ to be on shell, giving
\begin{equation}
 {\Delta}(x,z)^2
= (2-x-z) m_\varphi^2 + (x^2+z^2) m_\chi^2
> 0 \,.
\label{eq:Delta_hat_i}
\end{equation}

If we now perform the variable substitution $\alpha  = \lambda  - z(1-z)\beta$, then we once again reach an integrated product of two $\mathcal I$ functions up to two caveats: (i) the $\lambda$ integral has a lower bound of $z(1-z) \beta$ rather than $0$ and (ii) the integrand contains the phase factor arising from the product $p_E \cdot q_E$. But allowing the $\lambda$ integral to run from $0$ to $\infty$ and quenching the phase factor can only increase the value of the integral so that we reach the upper bound
\begin{align}
\delta \mathcal{M}^{(i)}_L ( \pmb \xi ,  \textbf 0)
&\leq
g^2 g_\chi^2 \sum_{\{ \textbf{n}, \textbf {m} \} \neq \textbf 0}
\int_{0}^{1} dx 
\int_{0}^{1-x} dz \,
\int_{q_E,k_E}
e^{i \textbf{q}\cdot  \ximn^{(i)}}
e^{i L \textbf{k}\cdot  \textbf m}  
 \int_{0}^{\infty} d\beta \, \beta^2 \,
e^{-\beta(k_E^2+ {\Delta}^2)}
\int_{0}^{\infty} d\lambda \,
e^{-\lambda(q_E^2 + m_\chi^2 )}  \,, \\
 &\leq
2 g^2 g_\chi^2 \sum_{\{ \textbf{n}, \textbf {m} \} \neq \textbf 0}
\int_{0}^{1} dx 
\int_{0}^{1-x} dz \,
\mathcal{I}_1 \big [  \vert \ximn^{(i)} (x,z)   \vert ; m_\chi \big ] 
\mathcal{I}_3 \big [  \vert L \textbf m    \vert ;  {\Delta}(x,z) \big ]  \,.
\end{align}
Exactly as with Diagram (j), we find that the $\xi$ dependence only appears in the $m_\chi$ integral and therefore that any enhancement due to the nonlocality of the operator is suppressed in this diagram by the heavier particle mass.

\bibliography{FV_non_Local} 

\end{document}